# Prediction of Radiotherapy-Induced Hematologic Toxicity in Cervical Cancer with Cohort-Aware Framework

Haotian Feng[1,*], Ph.D., Yan Kong[1,2,*], Ph.D., Tiantian Yang[2], M.S., Ke Sheng[1], Ph.D.

[1]Dept. of Radiation Oncology, University of California-San Francisco, San Francisco, United States
[2]Dept. of Radiation Oncology, Affiliated Hospital of Jiangnan University, Wuxi, Jiangsu, China
* These authors have equal contributions.

Corresponding author: Ke Sheng, Ke.Sheng@ucsf.edu
Statistical analysis author: Haotian Feng, Haotian.Feng@ucsf.edu

## ABSTRACT

**Purpose:** Hematologic toxicity (HT) remains a major dose-limiting complication for patients undergoing pelvic radiotherapy for cervical cancer. Standard dosimetric parameters are correlated with HT but show heterogeneous predictive accuracies. Imaging biomarkers from pelvic bone regions have demonstrated an additive predictive value by capturing the patient-specific hematopoietic reserve. However, unavoidable variability in the major bone segmentations degrades the model generalizability. To address this, we introduce a novel, cohort-aware representation-learning framework.

**Methods:** We retrospectively analyzed 152 cervical cancer patients who underwent pelvic radiotherapy without concurrent chemotherapy. Patients were divided into two distinct cohorts based on the operators performing pelvic bone segmentation. HT prediction models were developed across four methodologies: (1) cohort-specific training, (2) pooled training, (3) feature harmonization, and (4) our proposed cohort-aware neural network (CANN). The CANN architecture is uniquely designed to learn both shared and cohort-specific representations through contrastive regularization. Model performance was assessed via stratified cross-validation and an independent test set, with the area under the receiver operating characteristic curve (AUC) as primary evaluation metric.

**Results:** Cohort-specific models yielded cross-validation AUCs of 0.81 and 0.77 and test AUCs of 0.77 and 0.71 for Cohorts A and B, respectively. These results substantially outperformed the dosimetry-only model (test AUC=0.58). Pooling data across cohorts directly degraded predictive performance (cross-validation AUC=0.69; test AUC=0.64). While statistical harmonization provided only marginal gains, adversarial and correlation-based alignment methods exacerbated the performance decline, indicating negative transfer between the disparately contoured cohorts. Conversely, CANN robustly outperformed baseline approaches (cross-validation AUC=0.77 and test AUC=0.72). Ablation studies corroborated the critical role of both cohort-specific representations and contrastive alignment in achieving this robustness.

**Conclusions:** Including radiomic and dosiomic features significantly improved HT prediction compared with dosimetric metrics alone, but they also introduce sensitivity to contour variability.

This generalizability challenge of radiomics and dosiomics was effectively mitigated using a cohort-aware representation-learning approach.

# 1. Introduction

Hematologic toxicity (HT) remains a common and clinically significant adverse effect in cervical cancer patients undergoing radiotherapy(1). Pelvic bone marrow (BM), distributed across the ilium, sacrum, proximal femur, and lower lumbar spine, provides essential hematopoietic reserve yet is highly radiosensitive(2). Radiation-induced BM injury impairs hematopoietic stem cell function and disrupts the microenvironment required for effective hematopoiesis, resulting in leukopenia, anemia, and thrombocytopenia(3). Considerable interindividual variability in BM radiosensitivity and hematopoietic reserve further complicates HT risk stratification, underscoring the need for reliable, patient-specific prediction models(4), which can help guide personalized RT planning that minimizes the risk of HT.

Conventional HT prediction models rely on dosimetric and clinical parameters(5). The dose–volume histogram (DVH) metrics of pelvic BM have been widely studied, with low- to intermediate-dose exposure associated with HT risk(6). Clinical factors such as age, baseline blood counts, and treatment characteristics further improve prediction(7). However, these models are unable to capture patient-specific and spatially heterogeneous hematopoietic reserve (8, 9).

Radiomics enables noninvasive quantification of tissue characteristics from routinely acquired medical images(10). By capturing intensity and spatial heterogeneity, radiomic features derived from planning CT may reflect BM structural heterogeneity and radiation reserve(11). Prior studies indicate the efficacy of radiomics for predicting treatment response and toxicity, suggesting its potential for individualized HT risk assessment(12, 13).

Despite this potential, clinical translation remains challenging due to the sensitivity of radiomic features to segmentation variability(14, 15). Intraclass correlation coefficient (ICC) filtering(16) does not address systematic differences across cohorts. Feature harmonization methods like ComBat(17) attempt to align feature distributions and assume that cohort effects are noise, which can be removed without affecting predictive signals. In practice, cohort differences often reflect non-stochastic drift, including changes in imaging protocols, treatment techniques, and contouring practices(18, 19).

To address this limitation, we propose a **cohort-aware neural network (CANN)** that models both shared and cohort-specific representations. Rather than enforcing global feature alignment, the model captures common toxicity-related patterns while preserving cohort-specific variability, enabling more robust integration of heterogeneous datasets.

# 2. Methods

This study adhered to the TRIPOD statement(20) to ensure transparent reporting of model development, validation, and performance assessment. The complete TRIPOD checklist is provided in the Supplementary Materials.

## 2.1 Dataset

### Patients

A total of 203 patients were collected in this study, and 51 patients were excluded due to incomplete clinical information. Thus, this retrospective study included 152 cervical cancer patients (Cohort A: n=63; Cohort B: n=89) treated with pelvic EBRT at the Affiliated Hospital of Jiangnan University (2021–2023). The two cohorts were separated, with pelvic bone volume delineations performed at different time points by two different radiation oncologists. As such, variations between cohorts reflect systematic inter-observer and temporal practice differences in segmentation, rather than random variation, which could consequently introduce variability in image-based radiomic features(21).

To isolate radiation-induced hematologic toxicity (HT) and minimize systemic confounding factors, only patients receiving radiotherapy alone were included. Eligibility required pathologically confirmed cervical cancer, no prior pelvic RT or distant metastasis, and complete CBC data (pre-, during, and one-month post-RT). Patients with baseline grade ≥ 1 HT, long-term anemia, or secondary malignancies were excluded.

### Radiotherapy delivery

All patients underwent CT simulation and subsequent treatment in the supine position with a full bladder. CT scans were performed using a Light Speed Ultra scanner (GE Healthcare, USA) with the following parameters: slice thickness 5 mm, tube voltage 120 kVp, automatic tube current, and matrix size 512 × 512 pixels. EBRT was delivered using volumetric modulated arc therapy (VMAT), consisting of two coplanar mirrored arcs. Treatment planning was performed in the Varian Eclipse system (version 13.6) with the Acuros XB (AXB) algorithm, and the plans were delivered on Varian VitalBeam. The prescribed dose to the planning target volume (PTV) was 45Gy-50.4Gy in 25-28 fractions. BM was not delineated as an avoidance structure in any treatment plan.

### Hematologic toxicity evaluation

Complete blood count data, including white blood cell (WBC) count, absolute neutrophil count (ANC), hemoglobin (HGB), and platelet (PLT) levels, were collected from 1 week before the initiation of radiotherapy to 1 month after its completion(22). HT was graded according to the Common Terminology Criteria for Adverse Events (CTCAE), version 5.0(23). Acute HTs were defined as those occurring within 1 month after the completion of radiotherapy. A toxicity of Grade ≥ 2 in WBC, ANC, HGB, or PLT was defined as an acute HT event and was considered a positive endpoint in the subsequent analysis.

## 2.2 Feature Extraction

Pelvic bony structures were delineated as regions of interest (ROIs) on treatment planning CT images and used as a surrogate for pelvic bone marrow(24). The pelvis was defined as extending from the superior border of the L5 vertebral body to the inferior border of the ischial tuberosities. The pelvis was further divided into three subregions according to previously reported pelvic bone marrow partitioning methods(25): (1) iliac bone; (2) lower pelvis; and (3) lumbosacral vertebrae. A semi-automatic contouring workflow was adopted. Pelvic bones and femoral heads were first automatically segmented using a commercial auto-segmentation software (PVmed Co., Ltd., Guangzhou, China). The ROIs were then refined in 3D Slicer using Boolean operations and manual adjustments according to predefined anatomical boundaries by radiation oncologists. An example illustrating the relative locations of the ROIs is shown in **Appendix A**.

Three categories of features were extracted: radiomic/dosiomic, clinical, and dosimetric. Radiomic and dosiomic features were computed using PyRadiomics(26) from CT and dose images, including shape, first-order, and texture features (GLCM, GLDM, GLRLM, GLSZM, and NGTDM), extracted from both original and wavelet-transformed images.

Besides the radiomic features, clinical features, including body mass index (BMI), age, pathological type, FIGO stage(27), and complete blood count (white blood cell count, absolute neutrophil count, hemoglobin, and platelet count), were collected for all patients. Dosimetric parameters were extracted from each region, including $V5_{Gy}$, $V10_{Gy}$, $V20_{Gy}$, $V30_{Gy}$, $V35_{Gy}$, $V40_{Gy}$, $V45_{Gy}$, $V50_{Gy}$, as well as $D_{1\%}$, $D_{2\%}$, maximum dose, and mean dose, where Vx represents the normalized volume receiving at least x Gy, and $D_{x\%}$ represents the dose received by x percentage of the volume.

## 2.3 Feature Aggregation and Selection

We combined radiomic, dosiomic, dosimetric, and clinical features and reduced the feature dimension using the SelectKBest algorithm, a univariate feature selection method that ranks features based on their statistical relevance to the target variable. Specifically, we employed the Analysis of Variance (ANOVA) F-test as the scoring function to assess the variances between groups. Features with higher F-scores indicate stronger discriminatory power.

To determine the optimal feature dimensionality, we performed sensitivity analysis across multiple model architectures within the training folds for radiomic or dosiomic features. The number of selected features was varied, and model performance was evaluated using 5-fold cross-validation AUC. A subset of 100 features was chosen as it provided the best balance between predictive performance and model stability while minimizing overfitting (**Appendix B**).

## 2.4 Prediction Model

### Training and Evaluation

A five-fold cross-validation framework with train–validation–test splits was used (60%-20%-20%). Model performance was evaluated on testing set using AUC. All preprocessing steps were performed within training folds to prevent information leakage.

## Traditional Baselines Methods

To address cohort-related variability in radiomics, we evaluated several baseline approaches, including statistical harmonization and domain adaptation methods.

ICC Filtering(16): Features with an intraclass correlation coefficient (ICC) > 0.7 were retained to ensure reproducibility.

ComBat(17) and CovBat(28): ComBat corrects cohort-specific mean and variance differences using an empirical Bayes approach, while CovBat further aligns covariance structure to account for inter-feature correlations.

Domain-adversarial neural network (DANN)(29): DANN uses a gradient reversal layer to learn cohort-invariant representations, aiming to reduce domain-specific bias.

## Cohort-Aware Neural Network (CANN)

Assuming feature space $X \in \mathcal{R}^d$, label $Y \in [0,1]$ as well as cohort source $S \in [0,1]$. Classical harmonization like ComBat tries to remove any dependencies between X and S, however, for most medical data, this is usually not effective since the cohort and outcome are usually dependent due to common reasons like mask delineation, treatment protocol differences and scanner-protocol coupling. Our proposed CANN architecture models the outcome predictors conditioned on cohort information:

$$P(Y|X) = \sum_{s} P(Y|X, S = s)P(S = s|X)$$

The framework incorporates the cohort information into the prediction model to enhance the awareness of the patient cohort source more robust prediction, as shown in **Fig. 1**. CANN tries to address the limitations of these methods by introducing anatomical granularity into the harmonization process. CANN integrates cohort-aware priors (e.g., anatomical region labels) directly into the network layers. This allows the model to:

1. Apply differential weights to features based on their regional stability.
2. Preserve localized biological variance that global methods like CovBat might otherwise suppress.
3. Dynamically adapt the harmonization strategy to the specific skeletal cohort.

Two sub-frameworks of CANN are considered in this study: the CANN-Explicit (CANN-E) and CANN-Meta (CANN-M). CANN-E represented the Weighted Prediction Network as a function of cohort probability p:

$$Label = p * site_A + (1 - p) * site_B$$

While CANN-M used an MLP to predict label from the cohort model and probability.

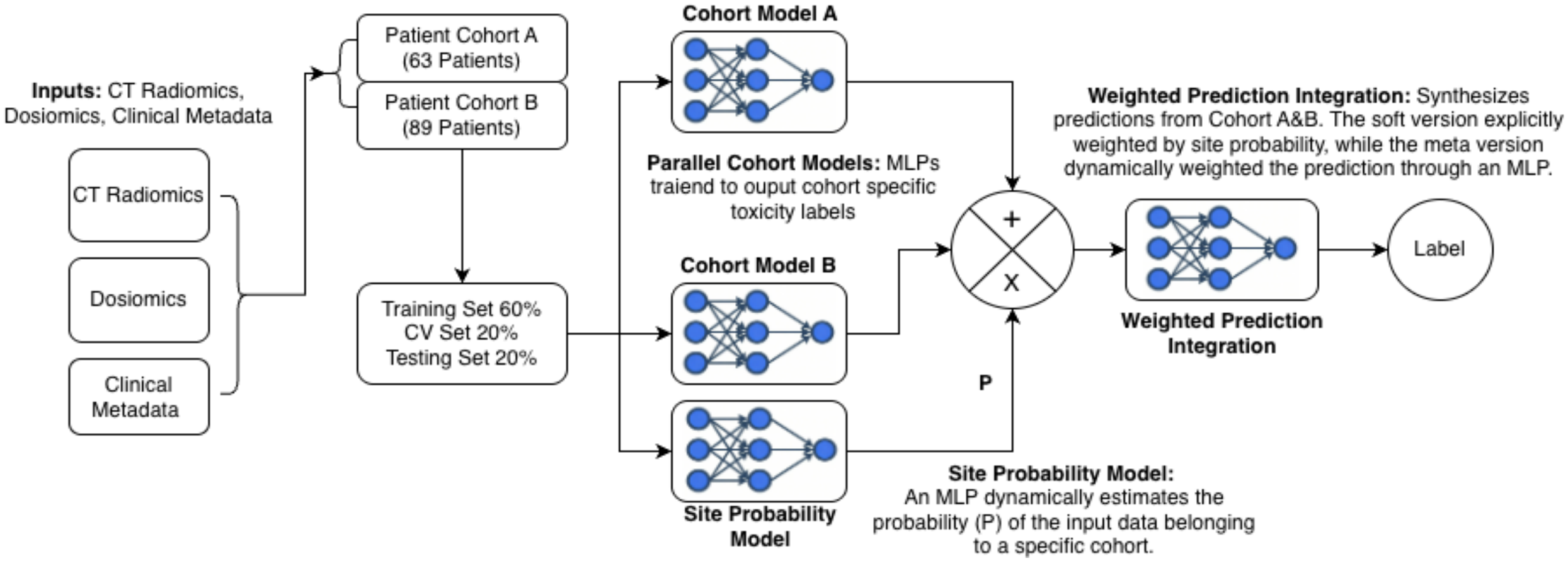

**Fig. 1.** Architecture of CANN Framework. The proposed framework is designed to manage feature shifts between multicentric patient cohorts (A and B) by conditioning outcome predictors on cohort-specific information. Features extracted from CT images, dose maps (dosimetry and dosiomics), and clinical metadata are processed through three parallel components: (1) Parallel Cohort Models: Individual Multi-Layer Perceptron (MLP) networks consisting of two hidden layers (64 neurons each) with ReLU activation, trained to output cohort-specific labels via Cohort Heads A & B. (2) Site Probability Model: A specialized MLP featuring two hidden layers (64 and 32 neurons, respectively) with ReLU activation, designed to estimate the Cohort Probability P for cohort A and B of the input data ($P_A$ and $P_B$), respectively. (3) Weighted Prediction Integration: A final integration layer that synthesizes the predictions from Cohort Heads A and B, weighted by the cohort probability P, to produce the final robust label (46% are positive labels (70 patients) and 54% are negative labels (82 patients)).

## 2.5 Model Interpretation Using SHAP

To interpret the contribution of individual features to model predictions, Shapley Additive Explanations (SHAP) were employed(30). SHAP values were calculated to quantify the impact of each feature on the predicted probability of hematologic toxicity. Feature importance was assessed based on the mean absolute SHAP value across the dataset.

# 3. Results

We first provided an overall comparison of predictive performance across different feature groups to establish the relative contribution of clinical, dosimetric, radiomic, and dosiomic features (Section 3.1). We then evaluated model performance within individual cohorts to characterize cohort-specific predictive behavior (Section 3.2), followed by cross-cohort analyses to assess generalizability and the impact of different harmonization strategies, including the proposed CANN (Section 3.3). To further investigate the underlying sources of performance variation, we examine cross-cohort differences in feature distributions and segmentation characteristics (Sections 3.4–3.5). Finally, we present model interpretability analyses to identify key predictive features and quantify the relative contributions of imaging, dosimetric, and clinical variables (Sections 3.6–3.7).

## 3.1 Overall model performance across feature groups

We first performed a high-level comparison of predictive performance across different feature groups, including dosimetric, clinical, radiomic, and dosiomic features. Models based solely on conventional dosimetric features demonstrated limited predictive ability, with AUC values around 0.52 and 0.58 for Cohort A and B individually, and an AUC of 0.50 for the combined value, indicating potential cohort-dependent discrepancy. The addition of clinical features yielded only marginal improvements. The detailed univariate analysis of clinical and dosimetric variables are provided in **Appendices C and D**, respectively, further support the limited predictive value of pure dosimetric or clinical features.

In contrast, models incorporating radiomic and dosiomic features showed substantially improved performance, consistently achieving AUC values exceeding 0.70 across multiple classifiers. The highest performance was observed when combining radiomics, dosiomics, dosimetric, and clinical features, indicating that imaging-derived features provide complementary and nonredundant information beyond traditional dose–volume metrics.

Across all experiments, radiomics- and dosiomics-based models outperformed dosimetry-only approaches by approximately 0.15–0.20 AUC. This trend was consistent across both cross-validation and independent testing sets, highlighting the robustness of imaging-derived features for hematologic toxicity (HT) prediction (**Fig. 2**).

## 3.2 Cohort-specific HT prediction

Evaluation within individual cohorts demonstrated strong predictive performance for models incorporating radiomic and dosiomic features. Cohort-specific models achieved cross-validation AUCs of 0.81 and 0.77, with corresponding independent test AUCs of 0.77 and 0.71 for Cohorts A and B, respectively. In contrast, dosimetry-only models remained limited in predictive capability (AUC = 0.58), underscoring the insufficiency of conventional dose–volume metrics alone.

Despite these gains, cohort-dependent discrepancies persisted. While combined dosiomic-radiomic models performed well in both cohorts, the relative ranking of classifiers and absolute AUC values differed between datasets. Notably, while pooling data (Cohort A+B) generally improved model stability, it did not consistently increase absolute AUC, especially for lower-capacity models such as gradient boosting. This confirms that naive pooling can obscure, rather than reconcile, cohort-specific predictive structures.

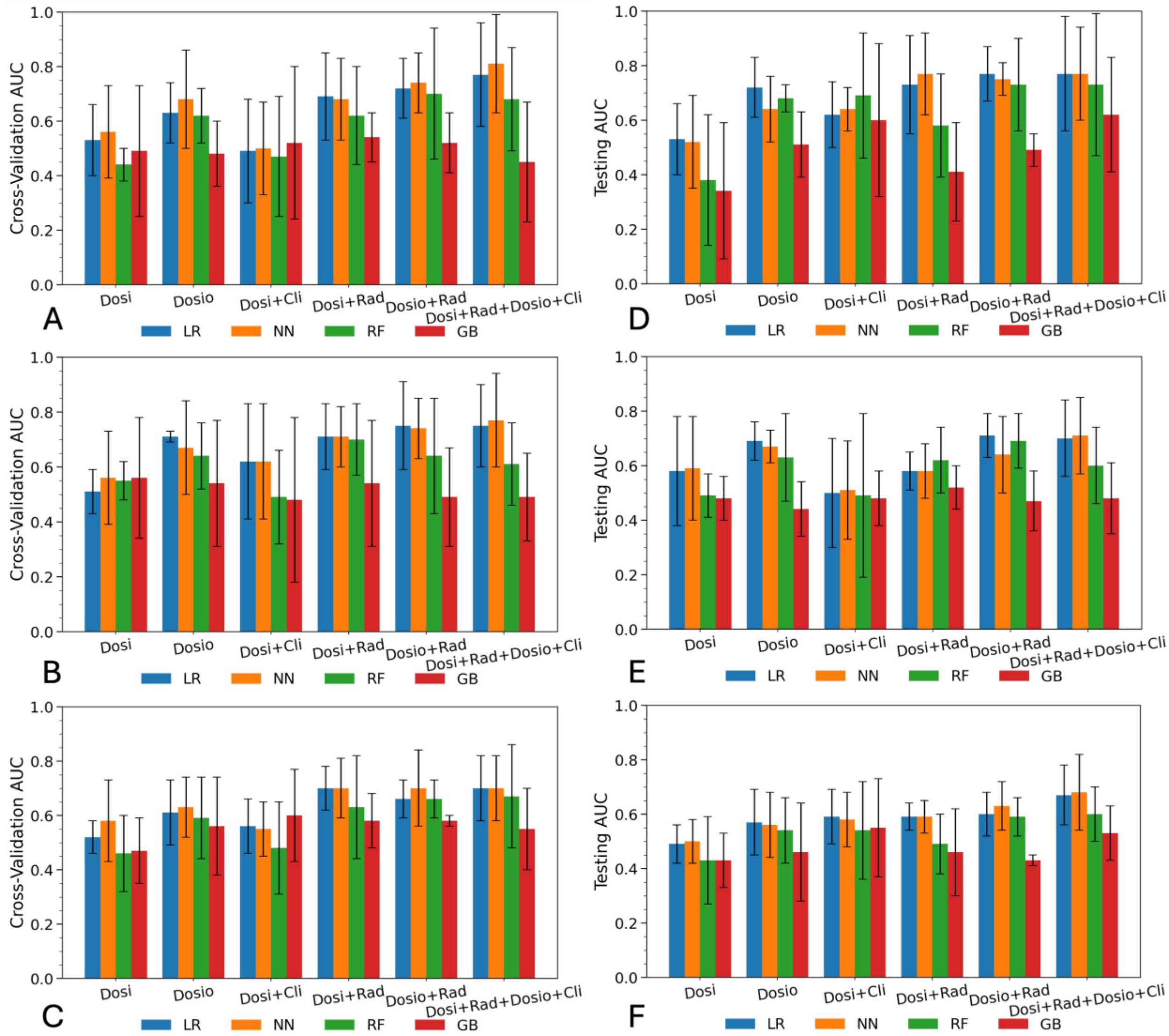

**Fig. 2.** Cohort specific HT prediction AUC (Mean, 95%CI; Dosi=Dosimetry, Dosio=Dosiomics, Rad=Radiomics, Cli=Clinical Features): (**A**) Cohort A cross-validation set prediction AUC; (**B**) Cohort B cross-validation set prediction AUC; (**C**) Cohort A&B cross-validation set prediction AUC; (**D**) Cohort A testing set prediction AUC; (**E**) Cohort B testing set prediction AUC; (**F**) Cohort A&B testing set prediction AUC.

## 3.3 Cross-cohort HT prediction

Models relying solely on dosimetry demonstrated limited cross-cohort generalizability compared to radiomics-derived representations (**Fig. 3A–B**). Among cohort-agnostic approaches, ComBat and CovBat provided only modest improvements, while DANN exhibited unstable performance, particularly with dosimetry-only inputs.

In comparison, the proposed CANN achieved the highest and most consistent performance across feature combinations. CANN-M attained a peak AUC of **0.77 (95% CI: [0.75, 0.80])** for all features setting in 5-fold cross-validation and **0.72 (95% CI: [0.63, 0.80])** for testing. The site probability model achieves an average AUC of 0.96 and an average accuracy of 0.95 for site label prediction.

These gains were most pronounced in high-dimensional imaging feature spaces, where cohort effects are complex and nonlinearly entangled. Permutation testing of cohort labels yielded no significant reduction in AUC ($p = 0.25$), confirming that CANN's improvements stem from modeling conditional cohort dependence rather than direct exploitation of cohort identity. This architecture ensures robust predictions that remain stable under cohort perturbations at inference time.

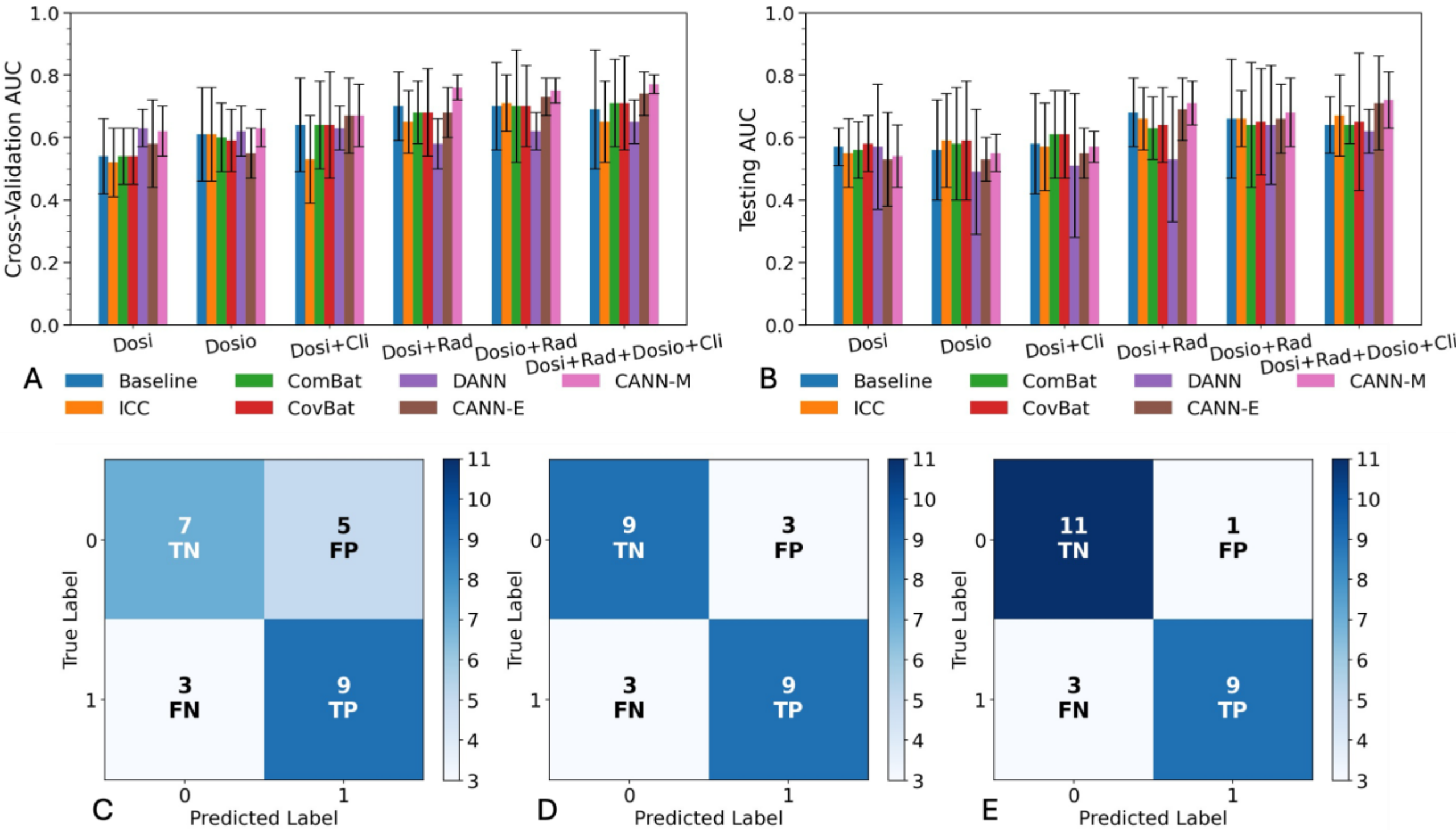

**Fig. 3.** Cross-Cohort HT Prediction AUC (Mean, 95%CI; Dosi=Dosimetry, Dosio=Dosiomics, Rad=Radiomics, Cli=Clinical Features): (**A**) cross-validation set prediction AUC; (**B**) testing set prediction AUC; (**C**) prediction confusion matrix from baseline pooled features; (**D**) prediction confusion matrix from ComBat harmonization; (**E**) prediction confusion matrix from CANN. TN denotes True Negative, FP denotes False Positive, FN denotes False Negative and TP denotes True Positive. CANN shows improved sensitivity to toxicity cases with a more balanced error profile across cohorts.

Beyond improved AUC, CANN demonstrated superior calibration and inter-cohort robustness. As shown in the confusion matrices for hematologic toxicity prediction (**Fig. 3C–E**), the baseline model (unharmonized) suffered from high misclassification rates and low sensitivity. While ComBat harmonization moderately improved classification balance, CANN achieved the most favorable error profile, significantly increasing true negative predictions and specificity while maintaining sensitivity. These results suggest that network-level cohort modeling allows for better adaptation to feature distribution shifts, providing more clinically relevant predictions than traditional pooling or harmonization methods. We further evaluated the clinical utility of these models using Decision Curve Analysis (DCA)(31) to determine the net clinical benefit against standard alternatives (**Appendix E**). Crucially, the proposed CANN model demonstrated strictly superior clinical net benefit compared to non-cohort-aware baseline model and standard harmonized model across the entire range of threshold probabilities (0.10 to 0.70).

## 3.4 Cross-cohort differences in radiomic feature distribution

We utilized Kolmogorov-Smirnov testing to reveal distinct regional variations in radiomic feature stability between Cohorts A and B. Significant distributional differences ($p<0.05$) were most prevalent in the lumbosacral vertebrae (22.35%) and lower pelvis (21.65%), while the iliac bone demonstrated greater stability (only 8.59% significant differences).

The non-uniform nature of these domain shifts was visualized using the GLSZM Zone Variance feature (**Fig. 4**). Cohorts showed marked separation in both original ($p = 0.0162$) and wavelet-filtered ($p = 0.0127$) spaces in the lower pelvis (**Fig. 2A, D**), while higher overlap in the iliac bone (**Fig. 2B, E**) and lumbosacral vertebrae (**Fig. 2C, F**). These findings highlighted the non-uniform nature of domain shifts across different skeletal regions. This regional heterogeneity provides a strong rationale for our CANN, which was designed to better adapt to these localized variations.

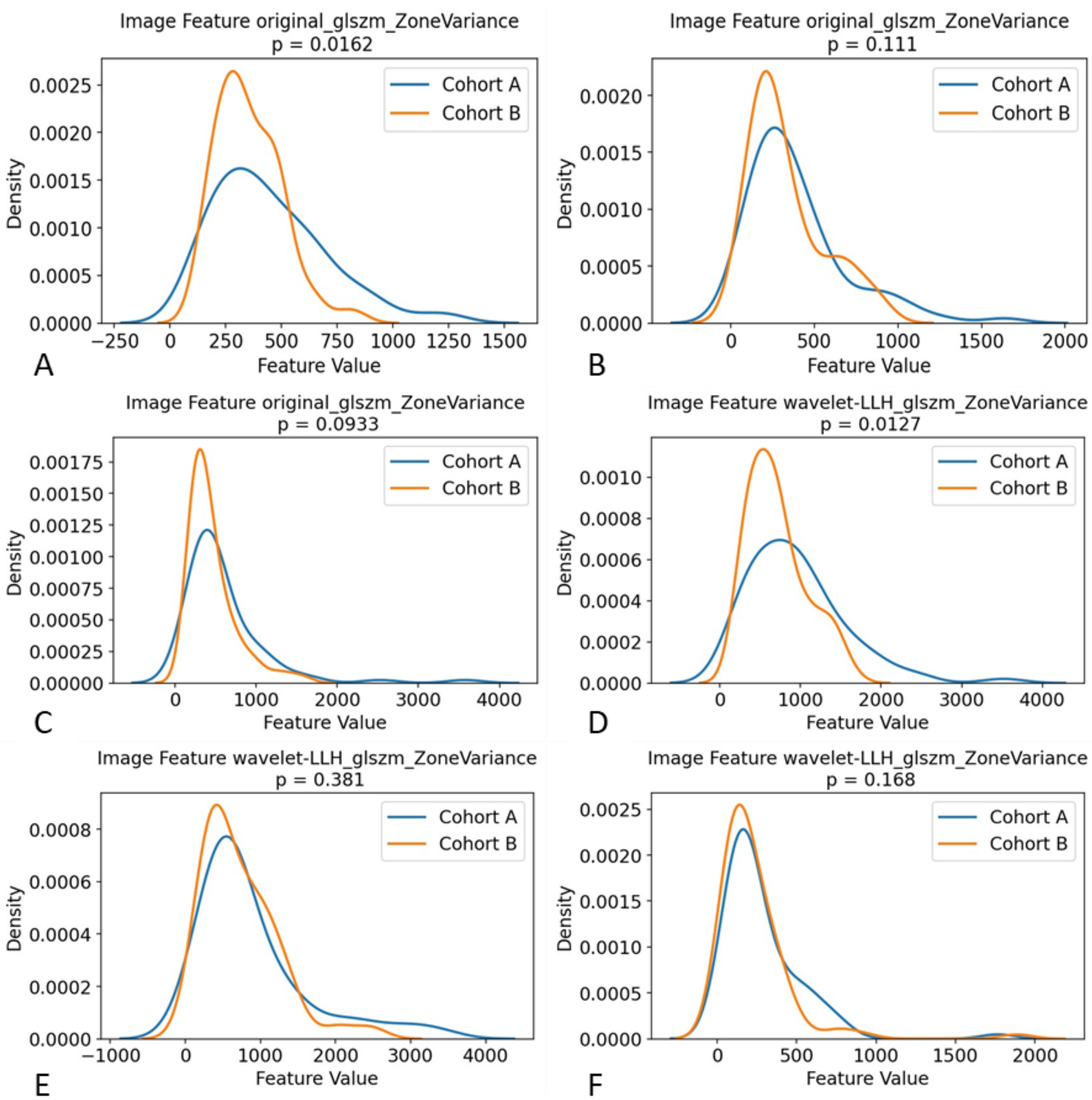

**Fig. 4.** Feature distributions between two patient cohorts for GLSZM Zone Variance in original space and one wavelet space at three bone regions: (**A,D**) lower pelvis; (**B,E**) iliac bone; (**C,F**) Lumbosacral Vertebrae.

## 3.5 Segmentation Differences Between Cohorts and Variability Analysis

Since both patient cohorts were imaged using the same scanner and reconstruction protocol, we focused on contouring consistency to investigate the origin of radiomic discrepancies. We observed a significant directional shift in lumbosacral volume, with Cohort B exhibiting lower volumes than Cohort A (mean difference ~ 18.68 cc). This discrepancy persisted after multivariable adjustment for height, age, and BMI ($p = 0.021$) and after height normalization ($p = 0.024$), indicating that the variance is not attributable to patient-related factors.

Visual inspection revealed a difference in delineation practices regarding the superior lumbosacral vertebrae (LSV) boundary (**Fig. 5A, B**). Cohort A included the volume up to the L6 inferior border, while Cohort B consistently terminated contours at the superior border of L5, as indicated by the red block. While both segmentation practices are considered acceptable, this slight discrepancy in the cranial-caudal margins introduces a significant distributional shift in the volume ($p = 7.516 \times 10^{-3}$, **Fig. 5C**). A similar significant volume distribution shift is also observed over the lower pelvis region ($p = 8.265 \times 10^{-3}$), while not in the iliac bone region ($p = 4.351 \times 10^{-1}$). Moreover, the masks in Cohort B are consistently smaller than Cohort A (lumbosacral vertebrae by 5.9%, lower pelvis by 5.7%, iliac bone by 1.9%).

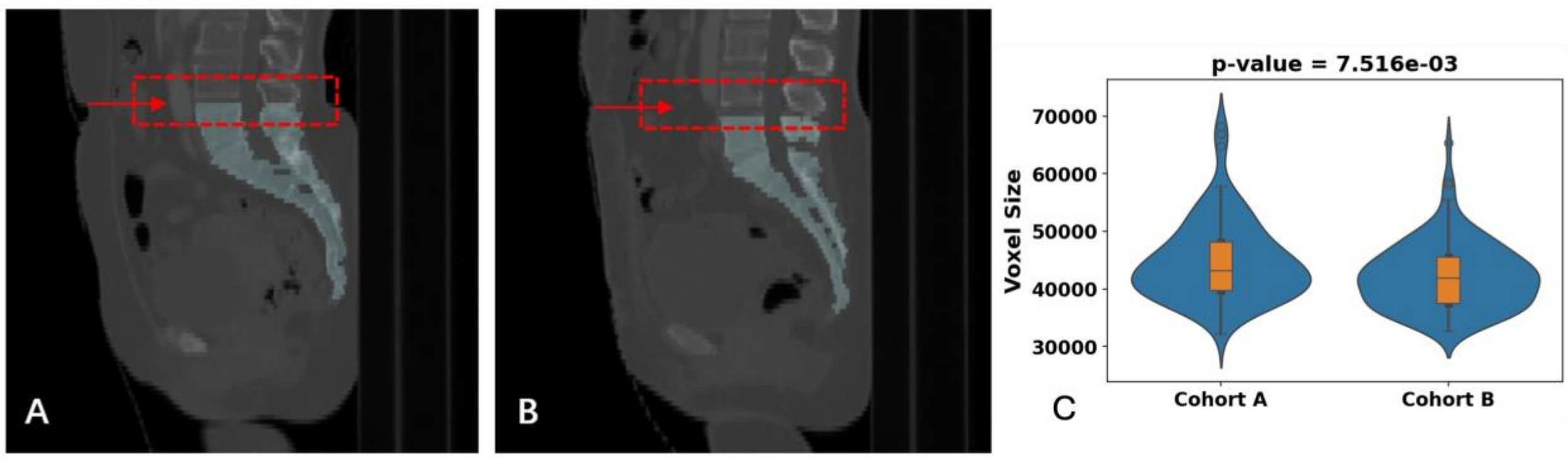

**Fig. 5**. Representative examples of lumbosacral vertebrae delineation over two cohorts: (A) cohort A; (B) cohort B; (C) Violin plot between the lumbosacral vertebrae mask over two cohorts.

## 3.6 Feature importance analysis using SHAP

SHAP analysis (**Fig. 6A**) highlights the top 20 predictors, integrating CT radiomics, dosiomics, dosimetry, and clinical variables in one representative fold. Overall, the model leveraged a heterogeneous set of features, reflecting the multifactorial nature of hematologic toxicity risk. Among clinical variables, baseline platelet and neutrophil counts, along with age, exhibited the highest absolute SHAP values; lower baseline counts were consistently associated with increased toxicity risk.

In the imaging domain, wavelet-transformed GLCM radiomic features contributed significantly, suggesting that bone marrow heterogeneity and spatial intensity patterns may encode biological susceptibility to radiation damage. Similarly, wavelet-based dosiomic metrics (GLSZM and GLRLM) captured spatial dose heterogeneity that exceeded the predictive value of conventional summary statistics. Notably, traditional dosimetry features, had a comparatively lower influence on individual predictions. Collectively, these results demonstrate that our model improves risk stratification by synthesizing complementary information from clinical status, imaging phenotypes, and spatial dose patterns.

Regionally, features derived from the lower pelvis accounted for the largest proportion of influential predictors, comprising an average of 34% of the top-ranked features, thereby highlighting the importance of this anatomic subregion in HT risk assessment.

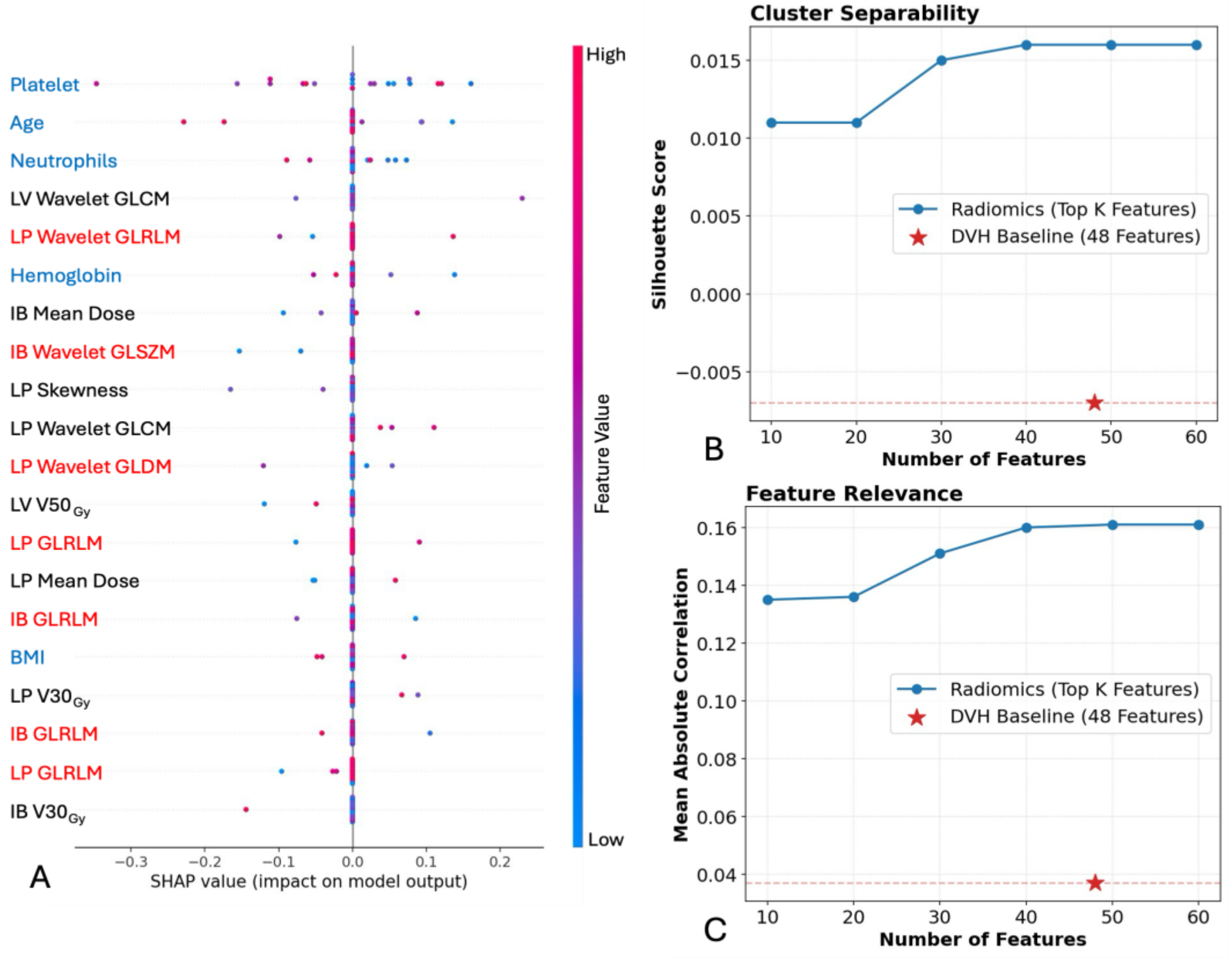

**Fig. 6**. LV denotes lumbosacral vertebrae, LP denotes lower pelvis and IB denotes iliac bone.
(**A**) SHAP summary plot illustrating feature importance and their impact on model predictions in one fold. Features are ranked by their average absolute SHAP values, with red and blue dots representing high and low feature values, respectively. Names of radiomic features are colored in black, dosiomic features are colored in red and clinical features are colored in blue. (**B**) Silhouette score between dosimetric parameters and different numbers of top combined features from SHAP. (**C**) Mean absolute correlation coefficient between dosimetric parameters and different numbers of top combined features from SHAP.

## 3.7 Interpretation of the dosimetric and radiomics contribution to HT prediction

The weak predictive signal of traditional dosimetry is evident in the split-violin plots of representative parameters, such as $V40_{Gy}$ and mean dose to the lower pelvis (**Appendix F**). Minimal separation between HT-positive and HT-negative distributions aligns with univariate results. This lack of divergence is partly due to high dosimetric consistency within the cohort; 52.1% of features showed a coefficient of variation (CoV) $< 0.1$. In contrast, dosiomic features exhibited significantly greater heterogeneity, preserving voxel-level spatial information that 1D dose-volume metrics lose.

Quantitative evaluations using the Silhouette Score (**Fig. 6B**) and Mean Absolute Correlation (Fig. 6C) also demonstrate the advantages of the combined features. While the dosimetric baseline results in near-zero correlation ($|r|=0.037$) and negative cluster separability ($S=-0.007$), the SHAP top features yield a four-fold increase in signal relevance ($|r|=0.161$) and a positive shift in cluster coherence. These findings suggest that texture-based descriptors from CT radiomics and dosiomics better reflect intrinsic susceptibility to hematologic toxicity than standard dosimetry, prioritizing nuanced microstructural and functional bone marrow variations.

# 4. Discussions

Accurate prediction of HT in cervical cancer patients undergoing radiotherapy can help personalize radiotherapy, including bone marrow-sparing RT(32) and proton therapy(33). Dosimetric features, which quantify radiation dose distribution within pelvic bone marrow subregions, have been widely reported as key predictors of treatment-related hematologic toxicity. For example, pelvic bone marrow $V10_{Gy}$ has been associated with Grade ≥2 HT, whereas $V30_{Gy}$ and $V40_{Gy}$ were not found to be significant in some cohorts(34). In contrast, Corbeau et al. reported $V10_{Gy}$, $V20_{Gy}$, and V40Gy as the significant predictors of HT(35), while Rahimy et al. identified $V20_{Gy}$ as strongly associated with HT(36). Konnerth et al. also reported a heterogeneous association of HT with low-dose vs. high-dose volumes(37). Similar heterogeneous dosimetric correlations have also been observed in rectal cancer patients, where conflicting associations between $V10_{Gy}$ and $V40_{Gy}$(38). The inconclusive dosimetric correlation across studies highlights a complex relationship between delivered dose and the incidence of HT. Our analysis shows that the dosimetric values of patients who experience HT do not differ significantly from those of patients without HT (**Fig. 3**). This observation was further confirmed by univariate analysis, which showed no statistically significant differences in $V5_{Gy}$–$V50_{Gy}$, mean dose, or maximum dose between HT-negative and HT-positive patients (**Appendix C**). This led to suboptimal HT prediction with dosimetric and clinical features alone. While functional imaging modalities (FLT-PET/MRI) can characterize marrow activity (39), they are not routinely available for treatment planning and are subject to variability in interpretation. In contrast, CT radiomics offers a more practical, readily integrated source of patient-specific data.

As shown by the current and previous studies, incorporating radiomic and dosiomic features substantially improved predictive performance, suggesting that imaging-derived features capture patient-specific biological characteristics beyond conventional dosimetric descriptors. Therefore, the model can be used to evaluate and guide personalized treatment plans that minimize HT. One potential challenge, as shown by the study, is limited model generalization. As shown by the study, a large proportion of radiomic features differed significantly between cohorts, which leads to substantial performance degradation across datasets.

A key source of this variation was identified as the delineation of pelvic bone marrow. While the iliac bone is relatively well defined, the lower pelvis and lumbosacral spine involve complex anatomical boundaries that are inherently prone to inter-observer variability. For example, the lumbosacral spine represents a transitional anatomical region, where identification of vertebral levels and superior–inferior boundaries can vary across observers. Similarly, the lower pelvis includes multiple irregular bony structures—such as the pubis, ischium, acetabulum, and proximal femur—further increasing delineation uncertainty. Prior studies have reported Dice similarity coefficients (DSC) of approximately 0.85 for a single cohort in cervical cancer treatment(40), while the value of DSC could decrease to 0.73~0.76 based on agreements among different radiation oncologists(41). Additional factors such as CT slice thickness, spatial resolution, and subtle variations in patient positioning during simulation may further affect boundary identification. The subtle yet systematic shifts are common among patients from different institutions and among observers. Auto-segmentation may help reduce inconsistency, but additional manual refinement is often necessary. In the current study, both cohorts used the same auto segmentation as the starting point, yet subsequent manual contour editing introduced a systematic shift in distribution, leading to variability in radiomics that cannot be eliminated(21).

Such cohort-dependent feature distributions present a fundamental challenge for radiomics-based predictive modeling. Conventional harmonization techniques rely on assumptions that cohort differences primarily reflect technical biases and that feature–outcome relationships are otherwise preserved across datasets. The developed CANN explicitly incorporates cohort information during training while avoiding dependence on cohort identity during inference. Rather than enforcing strict feature distribution alignment, the proposed framework learns shared predictive representations while allowing cohort-specific variations in feature expression. This design enables the model to preserve outcome-relevant signals that manifest differently across cohorts while mitigating the adverse effects of distribution shift. As a result, CANN demonstrated improved cross-cohort generalization compared with conventional harmonization approaches. Similar cohort-based frameworks have also been proven to be effective recently when dealing with lung cancer risk prediction(42) and neural representation learning for healthcare analytics(43).

The predominance of lower pelvic features in the SHAP analysis suggests that this subregion may play a disproportionate role in hematologic toxicity risk. The lower pelvis contains a substantial fraction of active bone marrow in adults and is frequently exposed to higher radiation doses during pelvic radiotherapy(44). It also aligns with prior studies demonstrating that dose to the lower pelvis is an independent predictor of hematologic toxicity(45, 46).

Several limitations should be acknowledged. First, while a homogeneous imaging protocol for both patient cohorts allowed us to focus on the impact of ROI segmentation, the same datasets also limited our ability to study model performance in relation to variations in the CT images. Additional validation using independent multi-institutional datasets will be necessary to assess generalizability. Second, representing cohort effects as a discrete variable does not fully capture finer-grained sources of variability, such as differences in patient characteristics and inter-observer contouring practices. Future work may explore hierarchical or continuous representations of cohort attributes to better capture these sources of variation.

# 5. Conclusions

Combining CT radiomics and 3D dosiomics significantly improves the prediction of hematologic toxicity in cervical cancer. To address multi-cohort variability in pelvic bone segmentation, our proposed Cohort-Aware Neural Network (CANN) adaptively models cohort origins rather than enforcing artificial uniformity. This approach preserves essential clinical dependencies, thereby improving model robustness and reliability in the face of unavoidable systematic drift in manual segmentation.

# Resource Availability

The entire framework can be found on: https://github.com/Isaac0047/HT_Site_Aware_NN. Requests for raw data required to reproduce the findings and further information should be directed to corresponding author.

# Appendix

## A. Mask delineation of different pelvic bone regions

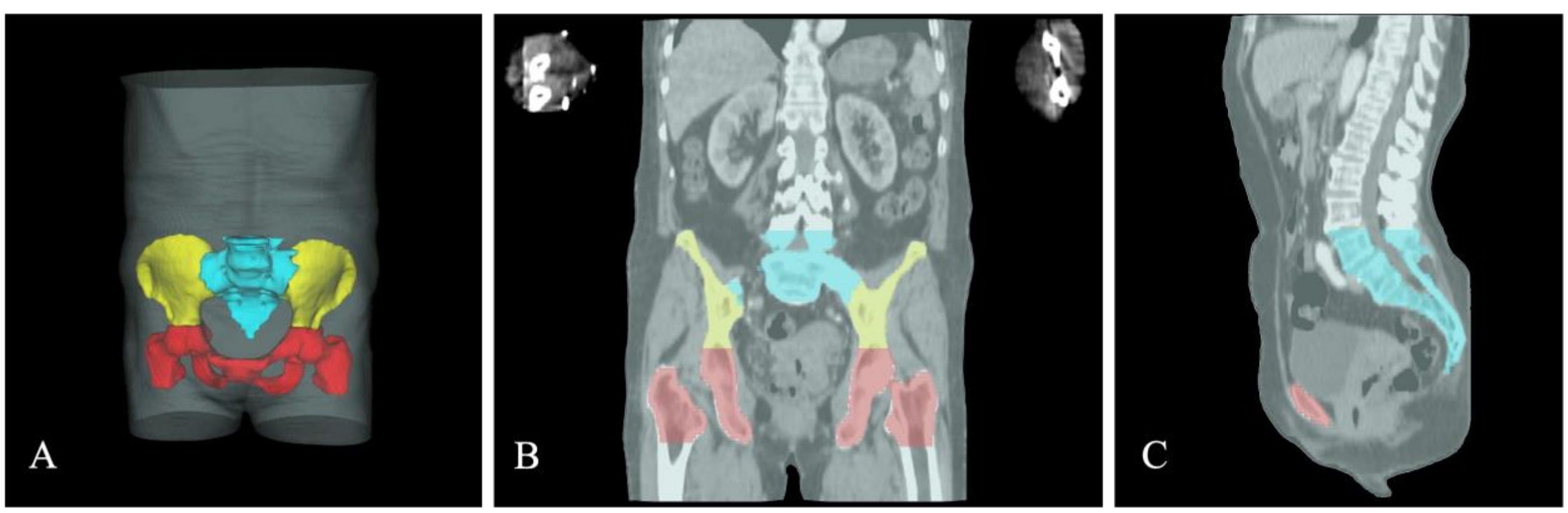

**Figure A1**. Spatial distribution of pelvic bone marrow regions. Lower pelvis (red), iliac bone (yellow), and lumbosacral vertebrae (blue) are shown relative to the body. (A) 3D rendering of the body and segmented bone marrow masks. (B) Coronal view of CT image with overlaid masks. (C) Sagittal view of CT image with overlaid masks.

## B. Prediction performance with respect to different number of features selected within cross-validation

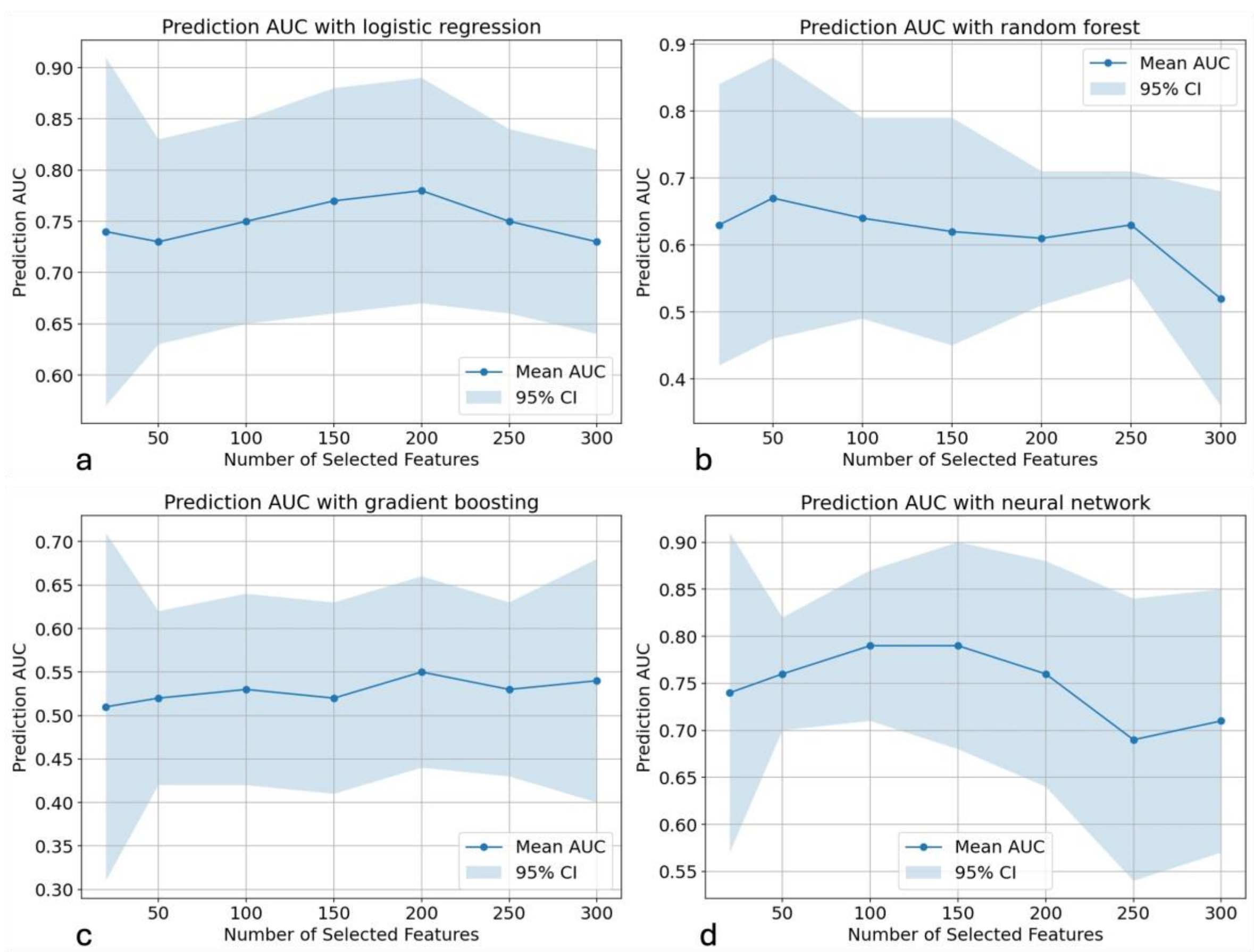

**Figure B1**. Prediction AUC with respect to different number of features (band denotes 95% confidence interval): (a) prediction AUC with logistic regression; (b) prediction AUC with random forest; (c) prediction AUC with gradient boosting; (d) prediction AUC with neural network.

## C. Univariate analysis of clinical features

Baseline characteristics between HT-negative and HT-positive patients were compared using Welch's t-test or Mann–Whitney U test for continuous variables and chi-square or Fisher's exact test for categorical variables, as appropriate. Baseline white blood cell count ($6.15 \pm 1.66$ vs $5.52 \pm 1.79 \times 10^9$/L, $p = 0.03$) and neutrophil count ($4.05 \pm 1.54$ vs $3.56 \pm 1.51 \times 10^9$/L, $p = 0.04$) were significantly lower in the HT-positive group. No significant differences were observed in age, BMI, red blood cell count, hemoglobin, platelet count, FIGO stage, or pathology type.

Table C1. Clinical features and hematologic toxicity analysis

| | | HT NEGATIVE | HT POSITIVE | P-VALUE |
|---|---|---|---|---|
| AGE | | 61.45 ± 11.72 | 58.49 ± 11.27 | 0.26 |
| BMI | | 23.24 ± 2.87 | 23.28 ± 3.27 | 0.96 |
| WHITE BLOOD CELL ($\times 10^9$/L) | | 6.15 ± 1.66 | 5.52 ± 1.79 | 0.03 |
| NEUROPHILS ($\times 10^9$/L) | | 4.05 ± 1.54 | 3.56 ± 1.51 | 0.04 |
| RED BLOOD CELL ($\times 10^9$/L) | | 4.05 ± 0.47 | 4.12 ± 0.47 | 0.40 |
| HEMOGLOBIN (G/L) | | 121.25 ± 11.26 | 122.69 ± 13.92 | 0.51 |
| PLATELET ($\times 10^9$/L) | | 231.82 ± 63.23 | 218.43 ± 65.00 | 0.23 |
| FIGO STAGE | 1 | 21 | 16 | 0.79 |
| | 2 | 31 | 23 | |
| | 3 | 26 | 26 | |
| | 4 | 4 | 5 | |
| PATHOLOGY TYPE | 1 | 74 | 62 | 0.48 |
| | 2 | 5 | 7 | |
| | 3 | 3 | 1 | |

## D. Univariate analysis of dosimetric features

Univariate analysis of dosimetric parameters across the lower pelvis, iliac bone, lumbosacral vertebrae, and the combined pelvic bone regions demonstrated largely comparable dosimetric distributions between the HT-negative and HT-positive groups. In the lower pelvis, most $Vx_{Gy}$ metrics and high-dose parameters (D1%, $D_{2\%}$, maximum dose) did not differ significantly between groups. Although the HT-positive cohort showed numerically higher exposure across multiple dose levels, only the mean dose reached nominal statistical significance.

Within the iliac bone region, no significant differences were observed for any dose–volume or point-dose parameters, with nearly identical mean values across groups. Similarly, for the lumbosacral vertebrae, all high- and intermediate-dose metrics were comparable between HT-negative and HT-positive patients. Although $V10_{Gy}$ demonstrated statistical significance ($p = 0.028$), the absolute difference between groups was minimal and occurred within a near-saturated dose range (approximately 100% volume coverage), suggesting limited clinical relevance.

Table D1. Dosimetric features and hematological toxicity analysis (p-value computed with t-test for approximately normal distribution and otherwise Mann-Whitney U test).

| REGION OF INTERESTS | DOSIMETRY | HT NEGATIVE | HT POSITIVE | P-VALUE |
|---|---|---|---|---|
| LOWER PELVIS | $V5_{Gy}$ | 0.92 ± 0.06 | 0.94 ± 0.05 | 5.33e-01 |
| | $V10_{Gy}$ | 0.81 ± 0.09 | 0.82 ± 0.09 | 6.00e-01 |
| | $V20_{Gy}$ | 0.54 ± 0.09 | 0.57 ± 0.08 | 1.84e-01 |
| | $V30_{Gy}$ | 0.30 ± 0.10 | 0.34 ± 0.12 | 1.51e-01 |
| | $V35_{Gy}$ | 0.21 ± 0.10 | 0.25 ± 0.12 | 2.41e-01 |
| | $V40_{Gy}$ | 0.14 ± 0.07 | 0.17 ± 0.09 | 1.78e-01 |
| | $V45_{Gy}$ | 0.07 ± 0.04 | 0.09 ± 0.04 | 5.16e-02 |
| | $V50_{Gy}$ | 0.01 ± 0.02 | 0.01 ± 0.01 | 4.46e-01 |
| | $D_{1p}$ | 49.39 ± 2.12 | 49.70 ± 1.85 | 3.23e-01 |
| | $D_{2p}$ | 48.59 ± 2.08 | 48.95 ± 1.80 | 2.45e-01 |
| | Max Dose (Gy) | 51.90 ± 2.41 | 52.38 ± 2.21 | 1.86e-01 |
| | Mean Dose (Gy) | 23.15 ± 3.28 | 24.13 ± 2.87 | 4.28e-02 |
| ILIAC BONE | $V5_{Gy}$ | 1.00 ± 0.00 | 1.00 ± 0.01 | 1.33e-01 |
| | $V10_{Gy}$ | 0.98 ± 0.02 | 0.98 ± 0.03 | 4.52e-01 |
| | $V20_{Gy}$ | 0.77 ± 0.10 | 0.79 ± 0.11 | 4.23e-01 |
| | $V30_{Gy}$ | 0.39 ± 0.12 | 0.40 ± 0.14 | 7.20e-01 |
| | $V35_{Gy}$ | 0.25 ± 0.09 | 0.27 ± 0.11 | 4.89e-01 |
| | $V40_{Gy}$ | 0.16 ± 0.06 | 0.17 ± 0.08 | 9.88e-01 |
| | $V45_{Gy}$ | 0.08 ± 0.04 | 0.09 ± 0.04 | 8.95e-01 |
| | $V50_{Gy}$ | 0.01 ± 0.02 | 0.01 ± 0.01 | 3.71e-01 |
| | $D_{1p}$ | 49.59 ± 1.80 | 49.87 ± 1.63 | 3.00e-01 |
| | $D_{2p}$ | 48.93 ± 1.81 | 49.18 ± 1.62 | 3.71e-01 |
| | Max Dose (Gy) | 51.65 ± 2.11 | 52.10 ± 2.18 | 1.82e-01 |
| | Mean Dose (Gy) | 28.31 ± 2.72 | 28.53 ± 2.79 | 6.21e-01 |
| LUMBOSACRAL VERTEBRAE | $V5_{Gy}$ | 1.00 ± 0.00 | 1.00 ± 0.00 | 5.84e-01 |
| | $V10_{Gy}$ | 1.00 ± 0.01 | 1.00 ± 0.00 | 2.81e-02 |
| | $V20_{Gy}$ | 0.95 ± 0.05 | 0.96 ± 0.03 | 1.62e-01 |
| | $V30_{Gy}$ | 0.70 ± 0.13 | 0.71 ± 0.12 | 6.85e-01 |
| | $V35_{Gy}$ | 0.54 ± 0.14 | 0.56 ± 0.13 | 6.34e-01 |
| | $V40_{Gy}$ | 0.40 ± 0.13 | 0.40 ± 0.11 | 7.51e-01 |
| | $V45_{Gy}$ | 0.24 ± 0.11 | 0.24 ± 0.09 | 1.00e+00 |
| | $V50_{Gy}$ | 0.05 ± 0.06 | 0.05 ± 0.06 | 5.49e-01 |
| | $D_{1p}$ | 50.15 ± 1.96 | 50.42 ± 1.84 | 3.68e-01 |
| | $D_{2p}$ | 49.83 ± 1.96 | 50.07 ± 1.86 | 4.18e-01 |
| | Max Dose (Gy) | 51.51 ± 2.23 | 51.78 ± 2.11 | 4.25e-01 |
| | Mean Dose (Gy) | 36.69 ± 3.26 | 36.95 ± 2.70 | 5.78e-01 |
| ALL REGION | $V5_{Gy}$ | 0.97 ± 0.02 | 0.98 ± 0.02 | 5.93e-01 |
| | $V10_{Gy}$ | 0.92 ± 0.04 | 0.93 ± 0.03 | 4.15e-01 |
| | $V20_{Gy}$ | 0.73 ± 0.05 | 0.75 ± 0.06 | 8.89e-02 |
| | $V30_{Gy}$ | 0.44 ± 0.09 | 0.46 ± 0.11 | 3.23e-01 |
| | $V35_{Gy}$ | 0.32 ± 0.08 | 0.34 ± 0.11 | 2.73e-01 |
| | $V40_{Gy}$ | 0.22 ± 0.06 | 0.23 ± 0.08 | 4.89e-01 |
| | $V45_{Gy}$ | 0.12 ± 0.04 | 0.13 ± 0.04 | 3.89e-01 |
| | $V50_{Gy}$ | 0.02 ± 0.03 | 0.02 ± 0.02 | 5.11e-01 |
| | $D_{1p}$ | 49.92 ± 1.92 | 50.23 ± 1.71 | 2.70e-01 |
| | $D_{2p}$ | 49.44 ± 1.89 | 49.70 ± 1.73 | 3.62e-01 |
| | Max Dose (Gy) | 52.17 ± 2.32 | 52.74 ± 2.15 | 1.07e-01 |
| | Mean Dose (Gy) | 28.66 ± 2.32 | 29.13 ± 2.19 | 1.86e-01 |

## E. Decision Curve Analysis (DCA) between Baseline, ComBat and CANN models

The large gap in net benefit between the blue CANN curve and baseline/ComBat curve in **Fig. E1** (especially between the clinically relevant region 0.4 to 0.6) highlights the superior in decision-making of the proposed CANN model. After threshold probability of 0.6, the baseline model's utility drops below 'Treat None', while CANN model remains stable and positive across the entire analyzed spectrum. Here, 'Treat None' corresponds to a universally non-interventional strategy while 'Treat All' corresponds to the opposite. The clinical utility function Net Benefit is defined as:

$$\mathrm{NB}(\mathrm{p_t}) = \mathrm{TPR} - \mathrm{FPR} \cdot \frac{\mathrm{p_t}}{1 - \mathrm{p_t}}$$

Where TPR represents True Positive Rate, FPR represents False Positive Rate, and $\mathrm{p_t}$ represents the threshold probability.

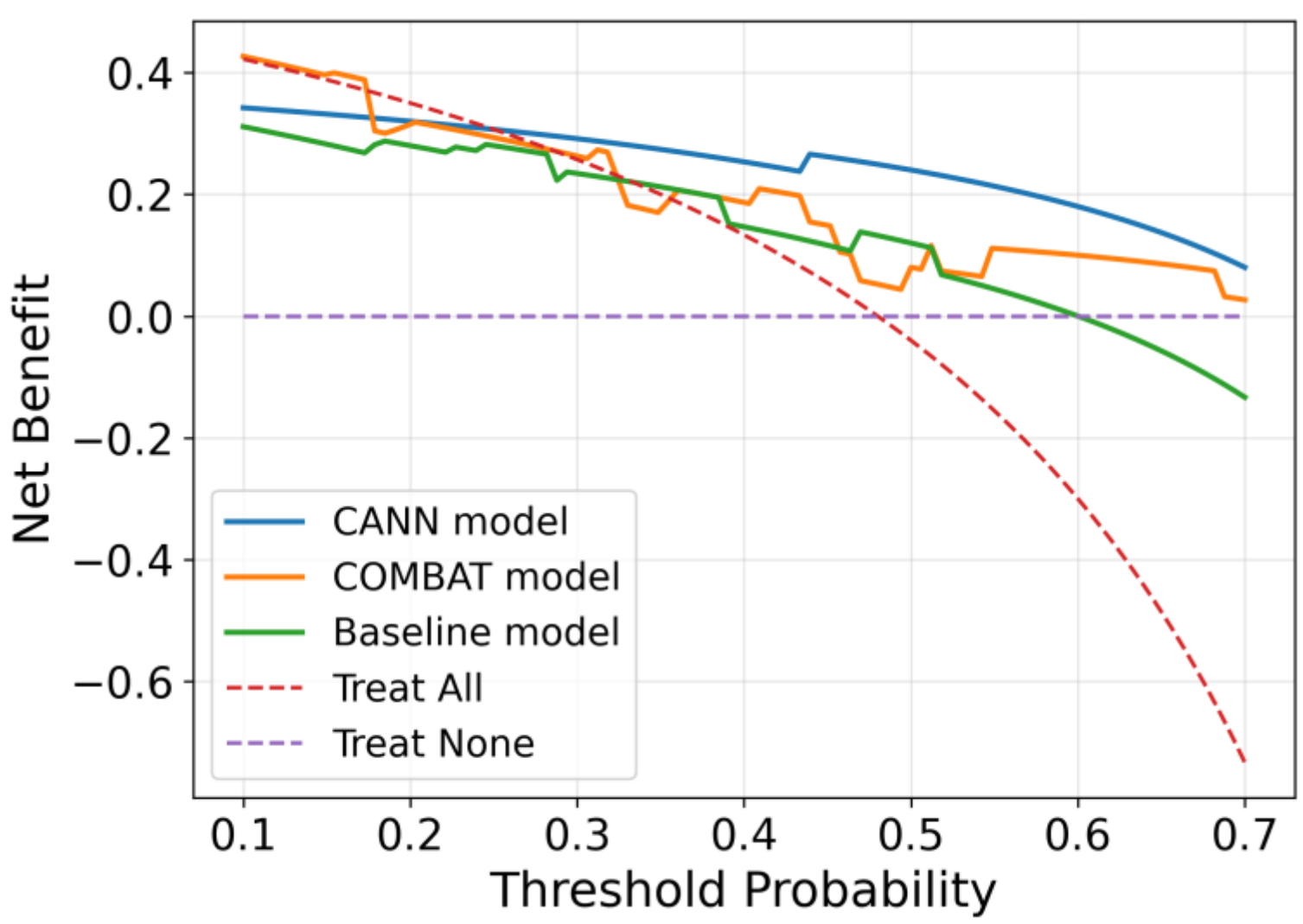

**Figure E1**. Decision Curve Analysis (DCA) for clinical net benefit comparison. The clinical utility of the CANN model (blue line) is compared against ComBat-harmonized model (orange line), baseline model (green line), theoretical strategies of 'Treat All' (red dashed line) and 'Treat None' (purple dashed line). The threshold probability is considered in the range between 0.1 to 0.7.

## F. Distribution of dosimetric features between HT-Negative and HT-Positive cohorts

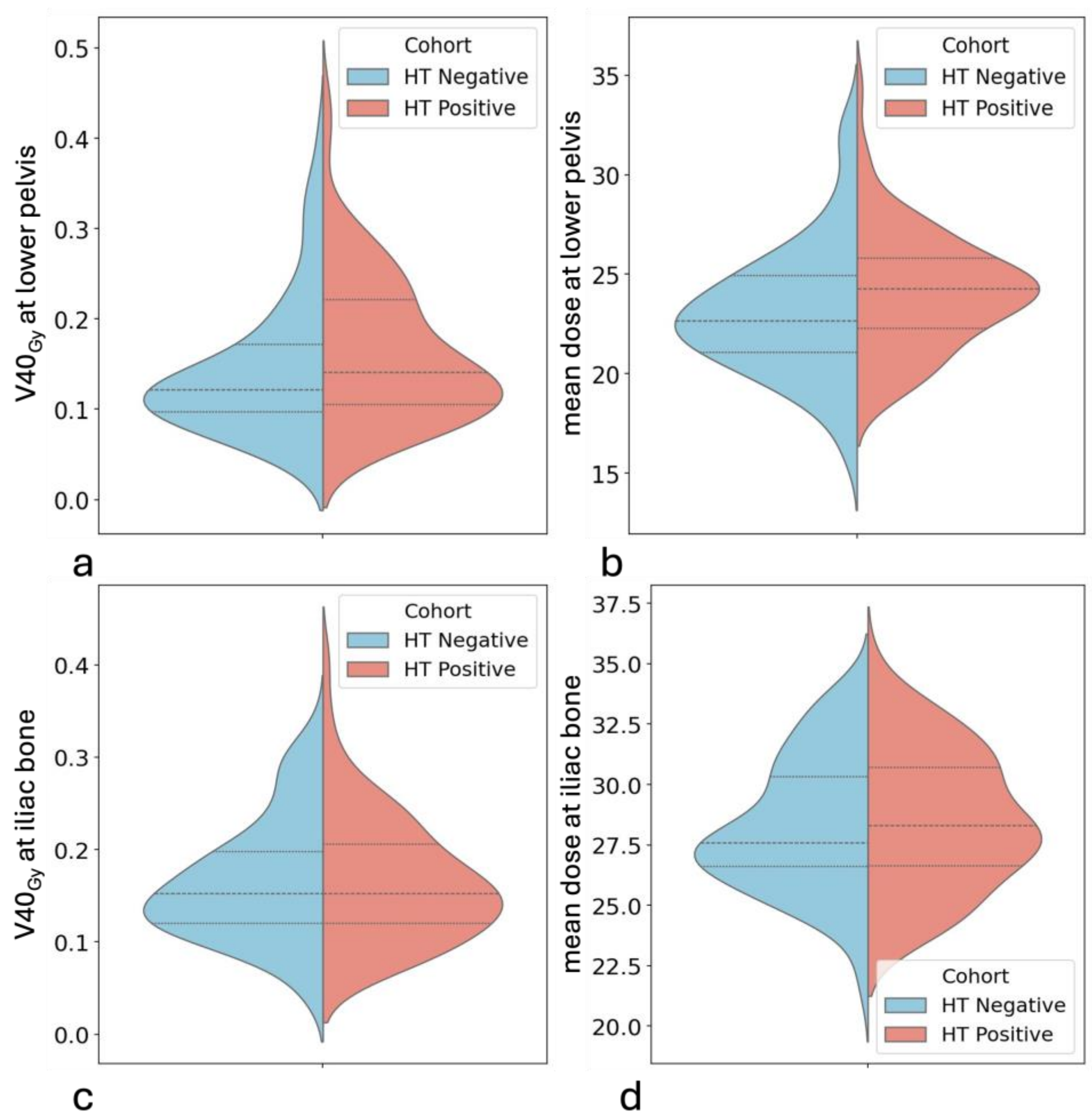

**Figure F1**. Comparative distribution of dosimetry features between HT-Negative and HT-Positive Cohorts: (a) $V40_{Gy}$ at the lower pelvis ($p=0.18$); (b) mean dose at the lower pelvis ($p=0.04$); (c) $V40_{Gy}$ at the iliac bone ($p=0.99$); (4) mean dose at the iliac bone ($p=0.62$). The thick dashed line represents the median value, and the two finer dotted line (upper and lower side) represent the inner quartiles ($25^{th}$ and $75^{th}$ percentiles).